\documentclass[apjl]{emulateapj}
\usepackage{color}

\usepackage[colorlinks=true, allcolors=blue]{hyperref}

\usepackage{graphics}

\usepackage{multirow}

\begin{document}

\title{Astrometric Methods for Detecting Exomoons Orbiting Imaged Exoplanets:\\ P\lowercase{rospects for} D\lowercase{etecting} M\lowercase{oons} O\lowercase{rbiting} \lowercase{a} G\lowercase{iant} P\lowercase{lanet in} $\alpha$ C\lowercase{entauri} A\lowercase{'s} H\lowercase{abitable} Z\lowercase{one}}
\color{black}
\shorttitle{Astrometric Detection of Exomoons}
\shortauthors{Wagner et al.}

\author{
 Kevin Wagner,\altaffilmark{1} Ewan Douglas,\altaffilmark{1} Steve Ertel,\altaffilmark{1,2} Kyran Grattan,\altaffilmark{3} S. Pete Worden,\altaffilmark{3} Aniket Sanghi,\altaffilmark{4,$\dagger$} Billy Quarles,\altaffilmark{5} \& Charles Beichman\altaffilmark{4}}


\altaffiltext{1}{Department of Astronomy and Steward Observatory, University of Arizona, AZ, USA}
\altaffiltext{2}{Large Binocular Telescope Observatory, AZ, USA}
\altaffiltext{3}{The Breakthrough Prize Foundation, CA, USA}
\altaffiltext{4}{Cahill Center for Astronomy and Astrophysics, California Institute of Technology, 1200 E. California Boulevard, MC 249-17, Pasadena, CA 91125, USA}
\altaffiltext{5}{Department of Physics and Astronomy, East Texas A\&M University, TX, USA}

\altaffiltext{$\dagger$}{NSF Graduate Research Fellow}


\begin{abstract}

Nearby giant exoplanets offer an opportunity to search for moons (exomoons) orbiting them. Here, we present a simulation framework for investigating the possibilities of detecting exomoons via their astrometric signal in planet-to-star relative astrometry. We focus our simulations on $\alpha$ Centauri A, orbited by a hypothetical giant planet consistent with candidate detections in Very Large Telescope and James Webb Space Telescope observations. We consider a variety of observatory architectures capable of searching for exomoons, including upcoming facilities and also a hypothetical dedicated facility $-$ e.g., a purpose-built space telescope with diameter = 3m, central observing wavelength of 500 nm, and contrast-limited performance of $\sim$10$^{-9}$ in 1 hr observations. We find that such a facility would be capable of detecting $\sim$Earth-mass moons in a five year campaign, assuming a Saturn-mass planet. More generally, we simulate expected detection limits for a variety of levels of astrometric precision. We find that moons as small as $\sim$0.2 M$_\oplus$ on orbital periods of 4$-$30 days can be detected with astrometric precision of 0.1 mas and observing cadence of 1 hr over a five year campaign. Additionally, we find that a 39m ground-based telescope can detect Earth-sized exomoons orbiting the same hypothetical planet with a more modest observing cadence of one day. We discuss these results as motivation for a dedicated space observatory as well as a more detailed study of the physical parameters of a greater variety of star-planet-moon systems.

\end{abstract}

\section{Introduction}
In the Solar System, every planet beyond Venus is orbited by at least one moon, suggesting that moons may be a common outcome of the formation of planetary systems. Earth's moon is $\sim$1\% the mass of the Earth, and a quarter of its radius, currently orbiting at $\sim$60 Earth-radii. Jupiter has dozens of moons, the four largest of which were independently discovered by Galileo Galilei and Simon Marius (\citealt{Galilei1610, Marius1614}; see also \citealt{Pasachoff2015}), now known as the Galilean satellites. Ganymede, the largest of the Galileian satellites has a mass $\sim$2\% and radius $\sim$41\% of the Earth, orbiting at $\sim$15$\times$ Jupiter's radius. Titan, orbiting Saturn, is similar in mass and radius to Ganymede, and orbits at $\sim$20 Saturn-radii. Mercury is only $\sim$5\% of the Earth's mass, and thus if Ganymede or Titan orbited the sun, they may be considered planets rather than moons. Perhaps the most extreme example among well-known bodies in the Solar System is the dwarf planet Pluto and its moon Charon, which have a mass ratio of 12\% and a center of mass that is outside of either body, determined through astrometry \citep{Foust1997}. 

While thousands of exoplanets have been discovered to date, few exomoon candidates have been identified. Searches have been performed to identify both transit-timing variations of the planets due to orbiting moons (e.g., \citealt{Simon2007, Kipping2009}) as well as photometric transits of the moons themselves (e.g., \citealt{Teachey2018a,Teachey2018b}). The first reported exomoon candidate was found with microlensing \citep{Bennet2014} $-$ however, degeneracy in the host lens mass and distance precludes an exact determination of the system properties. The first exomoon candidate around a transiting planet, super-Jupiter Kepler 1625b, was reported by \cite{Teachey2018b}. The candidate moon's mass is $\sim$10$\times$ that of Earth and $\sim$4$\times$ its radius $-$ similar to Neptune. This exomoon candidate is much larger than the Solar System moons, but is rather similar to the mass and radius \textit{ratios} of the Earth-moon system. However, alternative (systematic) explanations for the signal also exist \citep{Heller2019,Kreidberg2019}. A second, similarly large exomoon candidate, Kepler 1708b-i, was recently reported with mass of $\sim$2.6 M$_{\oplus}$ and orbital semi-major axis $\sim$12 planetary-radii \citep{Kipping2022, Kipping2025}. Like Kepler 1625b-i, other explanations for the signal have also been proposed  \citep{Heller2024}. 

Direct approaches for detecting exomoons have also been proposed. In particular, tidally heated moons could be hotter and thus brighter than their host planets, enabling them to be readily detected but also easily confused with planetary emission \citep{Limbach2013}. Spectroastrometry, the differential measurement of the shift in position of the center of light in wavelengths for which the moon is brightest compared to those for which the planet is brightest, may disentangle these signals \citep{Agol2015}. Radial velocity \citep{Vanderburg2018,Ruffio2023}, polarimetry \citep{Michaelis2025}, and radio emission \citep{Noyola2014} are other alternative means of detecting exomoons that have been investigated in past studies.

Astrometry enables detecting unseen companions via their gravitational influence on a more readily observable body. Absolute astrometry has enabled prioritizing imaging searches for exoplanets and substellar companions by targeting systems with astrometric accelerations measured between the Gaia and Hipparcos missions (e.g., \citealt{Brandt2021, Currie2023, Franson2023}). For systems with an astrometric reference, such as binary stars, relative astrometry is possible and often more constraining on the presence of unseen companions than other methods. For instance, using relative astrometry from \textit{Hubble Space Telescope} (HST) observations covering two decades, \cite{Bond2015} establish sensitivity to planets of $\sim$5 M$_{Jup}$ orbiting within the Procyon binary. A dedicated space mission to enable relative astrometry of $\alpha$ Centauri and other nearby binary systems, TOLIMAN \citep{Tuthill2018}, is expected to achieve sub-microarcsecond precision $-$ sufficient for the detection of Earth-mass planets. 

\begin{figure*}[htpb]
\centering
\plotone{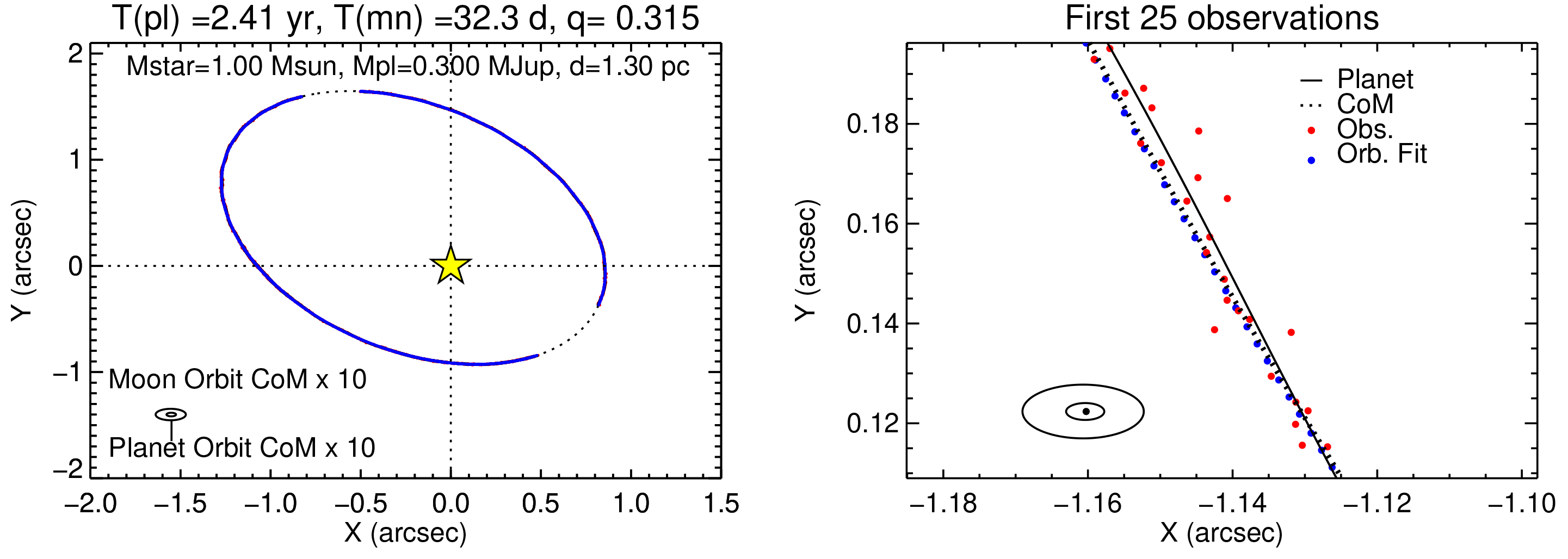}
\caption{\footnotesize \label{orbit} \textit{Left: }An illustrative example of a 30 M$_{\oplus}$ moon orbiting a Saturn-mass planet, with their center of mass (CoM) in a $\sim$1.8 au orbit around a 1.0 M$_{Sun}$ star at 1.3 pc (similar to the planet candidate around $\alpha$ Centuari A described in \citealt{Wagner2021,Beichman2025,Sanghi2025}). Such a large moon is not representative of moons that would be likely found in nature, but it serves to better illustrate several following points via its larger gravitational effect on the planet. The blue curve shows the orbit fit for the observed dates. The dashed curve shows the orbit of the center of mass (CoM) of the planet and moon. \textit{Right}: Same as left, but zoomed in. The moon and planet orbits about their CoM are shown at their actual scale, rather than enlarged as in left panel. While the orbit fit is performed for the measured planet position (red points), note that the orbit fit (blue points) most closely matches the position of the CoM (black dotted) $-$ illustrating that the moon's effect on the planet's position does not significantly affect the Keplerian orbit fit of the planet.}
\end{figure*}

\begin{figure*}[htpb]
\centering

\includegraphics[width=\textwidth]{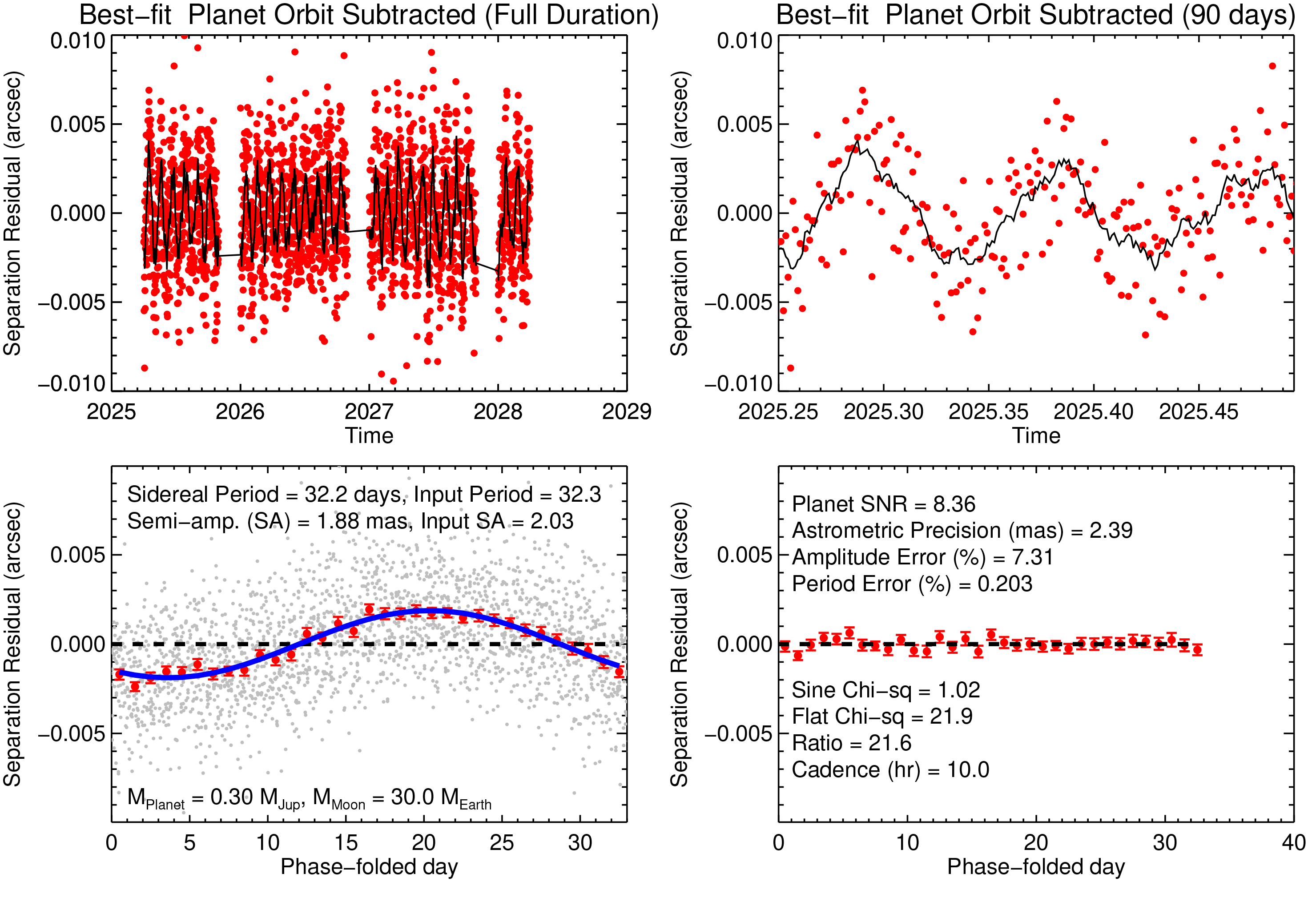}
\caption{\footnotesize \label{residuals} \textit{Top Left:} Separation residuals after subtracting the predicted positions of the planet based on a Keplerian orbit fit to the astrometry. The black curve shows a 10$\times$ smoothed version of the data. \textit{Top Right:} The same residuals, shown over a smaller range. \textit{Bottom Left:} Phase-folded separation residuals (gray points, binned by one day as red points) with maximum $\chi^2$ difference between the sine fit (blue curve) and null hypothesis (dashed line). \textit{Bottom Right:} Separation residuals with sine fit subtracted. The star-planet-moon system configuration corresponds to that in Fig. \ref{orbit}.}
\end{figure*}

A similar, but more challenging configuration for relative astrometry exists for imaged exoplanets, in which the relative position of the planet with respect to the star is readily measurable. In this work, we explore the question of whether exomoons orbiting nearby exoplanets could be detected with next generation ground$-$ and space-based telescopes. Some basic scaling relations are considered in \cite{Lazzoni2022}, which we summarize and expand on here. As the astrometric signal scales linearly with inverse distance, the closest systems will be the most promising $-$ should they host amenable planets. Moons around planets with larger radii will also be more readily detectable, as the astrometric precision scales with the inverse of the SNR, which (all else equal) scales approximately as the square of the radius of the planet, due to the dominant source of noise being from the star and/or thermal background. The astrometric signal is also directly proportional to the moon-to-planet mass ratio, and thus moons orbiting less massive planets (of a given radius) will also be more readily detectable. Other more complicated factors, such as how well the planetary orbit can be constrained, also play a significant role in determining exomoon detectability through relative astrometry. Therefore, the major goals of this study are to create a simulation framework to incorporate these combined effects into simple detectability metrics, and to recommend observing approaches and mission architectures to facilitate the detectability of nearby, potentially habitable exomoons within the next decade(s).

Finally, this study is further motivated by the habitable-zone giant planet candidate around $\alpha$ Centauri~A \citep{Wagner2021, Beichman2025, Sanghi2025}, which could mirror the Pandora-Polyphemus system in the popular science fiction series, \textit{Avatar} \citep{Cameron2009}. Very Large Telescope (VLT) observations  from 2019 \citep{Wagner2021} first detected a point source in the thermal infrared consistent with being a giant planet in $\alpha$ Centauri's habitable zone. More recently, James Webb Space Telescope (JWST) observations from 2024 \citep{Beichman2025,Sanghi2025} revealed a point source on the opposite side of the star. The two detections have positions and brightnesses that are consistent with orbital motion and a planetary atmosphere. However, the S/N of either detection is low, and thus to confidently confirm the planetary nature of one or both detections, a third detection is needed. The two existing detections result in a predicted location due to orbital motion to be tested with a third epoch (best observability in August 2026: see \citealt{Beichman2025}). If this planet were to be confirmed, it would be an ideal target for relative astrometric searches for exomoons with a distance of 1.3 pc, a planet-to-star separation of $\sim$1 arcsec, and a planet that is less massive than Jupiter \citep{Zhao2018}, but plausibly of similar radius \citep{Beichman2025}. 

Nearby habitable-zone planets, like the giant planet candidate in $\alpha$ Centauri, are not likely to transit their host stars and thus neither do their moons $-$ necessitating methods of detection such as relative astrometry, spectroastrometry (e.g., \citealt{Agol2015}), or eclipses (transits) of moons of their host planet \citep{Heller2016}. Here, we focus our simulations around $\alpha$ Centauri and its giant planet candidate, as a demonstration of the relative astrometry methodology and motivation for observatories under consideration. The physical parameters of all possible star-planet-moon system architectures constitute a complex parameter space that will be the subject of a future more detailed study. 

This paper is organized as follows: in \S\ref{simulation}, we describe our methodology for simulating exomoons in astrometric time series, as well as methodology for analyzing the simulated data to recover the moon signal and properties (period and amplitude). In \S\ref{results}, we describe our findings, which focus on the example case of exomoons orbiting a hypothetical giant planet in $\alpha$ Centauri's habitable zone (similar to the aforementioned candidate). In \S\ref{discuss}, we recount the method's limitations and assumptions, and discuss the applicability of the results to telescopes in construction or under consideration. In \S\ref{summary}, we briefly summarize our findings and conclusions. 

\section{Simulation Methodology }
\label{simulation}

\subsection{General System Parameterization}

Our aim is to develop a simple framework to extract the signal from an exomoon from a time-series of relative astrometry for an imaged exoplanet.\footnote{Our simulation software is publicly available at \url{https://github.com/astrowagner/exomoonsim}.} First, we begin by simulating the physical system, which includes the Keplerian orbit of the planet and moon's center of mass about the star, and that of the planet and moon about their center of mass. We assume a single exomoon, planetary orbital parameters, moon orbital parameters, and masses for each body, resulting in a series of true positions for each. Focusing on the case for which the moon's contribution to the planet+moon brightness is insignificant, we add Gaussian noise (measurement errors) to the simulated data with mean of zero and standard deviation equivalent to the optical system's full width at half maximum (FWHM, which depends on the observation wavelength and telescope diameter) divided by the signal to noise ratio (SNR), which itself depends on the assumed albedo, radius, orbital separation, and atmosphere of the planet, as well as the performance of the observatory (among other possible factors - e.g., rings).

We assume that measurement errors dominate the astrometric error budget, while in reality systematic terms such as detector deviations from isoplanism (distortion) will also play a role and will require calibration. Such systematic uncertainties are typically on the order of $\sim$mas (e.g., \citealt{Weible2025}). We also assume that SNR scales linearly with apparent planet brightness and with the square root of the duration of the observation (i.e., that the observations are not speckle-dominated). In \S \ref{results}, we simplify the Gaussian noise term with a single astrometric precision parameter, and discuss achieving this level of precision in \S \ref{discuss}. We also specify the observing cadence, amount of downtime between observations, and annual gaps in observability (i.e., assuming a target that is not within the observatory's continuous viewing zone).



\begin{deluxetable}{lllc}
\tablecaption{Exomoon System Simulation Parameters}
\tablewidth{0pt}
\tablehead{
\colhead{\textbf{Category}} & \colhead{\textbf{Parameter}} & \colhead{\textbf{Value}} & \colhead{\textbf{Description (Units)}}
}
\startdata
System & $d$ & 1.3 & Distance (pc) \\
                        & $M_{\star}$       & 1.0 & Star mass (M$_\odot$) \\
                         & $\tau_{ref}$         & 2000.0 & Ref. Epoch (yr) \\                      
\hline
Planet & $a_{pl}$        & 1.8 & Semi-major Axis (au) \\
                        &   $M_{pl}$   & 0.3 & Planet Mass (M$_{Jup}$) \\
                        & $i_{pl}$      & 45.0 & Planet Inclination ($^\circ$) \\
                        & $e_{pl}$      & 0.3 & Planet Eccentricity \\
                        & $\omega_{pl}$    & 150.0 & Arg. of Periapsis ($^\circ$) \\
                        & $\Omega_{pl}$ & 150.0 & Long. of Asc. Node ($^\circ$) \\
                        & $\tau_{0, pl}$      & 3.0 & Periastron Passage (yr) \\
\hline
Moon      & $a_{mn}$        & varied & Semi-major Axis (au) \\
                        &   $M_{mn}$   & varied & Moon Mass (M$_{\oplus}$) \\
                        & $i_{mn}$        & 50.0 & Moon Inclination ($^\circ$) \\
                        & $e_{mn}$        & 0.05 & Moon Eccentricity \\
                        & $\omega_{mn}$     & 0.0 & Arg. of Periapsis ($^\circ$) \\
                        & $\Omega_{mn}$   & 0.0 & Long. of Asc. Node ($^\circ$) \\
                        & $\tau_{0, mn}$         & 0.0 & Periastron Passage (yr) \\

\hline
Observation & start     & 2025.25 & Start year \\
                             & duration  & 1826.25 & Obs. Duration (days) \\
                             & $t_{exp}$      & varied & Indv. Exp. Time (hr) \\
                             & $t_{down}$    & varied & Downtime (hr) \\
                            & $\theta$    & varied & Sep. Precision (mas) \\
\hline
Period & $t_{bin}$ & 0.1 & Binning window (days) \\
Testing                               & $p_{start}$   & 4.0 & Start period (days) \\
                                & $p_{end}$     & 30.0 & End period (days) \\
                                & $p_{step}$ & 0.01 & Period step (days)\\
\enddata
\tablecomments{Static simulation parameters were used throughout this study, unless otherwise noted. Parameters were chosen to be similar to the $\alpha$ Centauri A system and its planet candidate \citep{Wagner2021,Beichman2025,Sanghi2025}. All inclinations are measured from the plane of the sky. \label{params}}
\end{deluxetable}

\subsection{Example of a 30 M$_\oplus$ Moon Orbiting a Candidate Giant Planet in $\alpha$ Centauri A's Habitable Zone}

We show examples of the simulation framework in Figs. \ref{orbit}-\ref{residuals} and list the parameters in Table \ref{params}. In these examples, we choose a relatively large and perhaps unlikely moon-to-planet mass ratio, to demonstrate the effects on the planet on scales that are readily identifiable on the plots. As we will show, the actual limits of current and future telescopes far surpass these illustrative examples. In particular, the examples in Figs. \ref{orbit}-\ref{residuals} include a planet mass of 0.3 M$_{Jup}$, and moon mass of 30 M$_{\oplus}$, for a mass ratio, $q\sim0.3$. We assume a star mass of 1.0 M$_{\odot}$ and distance of 1.3 pc, similar to $\alpha$ Centauri A. The planet's orbital parameters of semi-major axis, eccentricity, and inclination are ($a_{pl}$, $e_{pl}$, $i_{pl}$) = (1.8 au, 0.3, and 45$^\circ$), similar to the properties in \cite{Beichman2025} for the planet candidate. Inclinations, unless specified otherwise, are measured with respect to the plane of the sky. For the moon, in this example we use ($a_{mn}$, $e_{mn}$, $i_{mn}$) = (30 $R_{Jup}$, 0.05, and 50$^\circ$). This configuration is shown in Fig. \ref{orbit}. Other simulation parameters are listed in Table \ref{params}. Again, the example of such a massive moon on such a wide orbit is intended to be illustrative of the effects, and not of the limits of the method's sensitivity or a system that is likely to be found in nature (e.g., \citealt{Calibrisi2021} find that only 15\% of simulated satellite systems are more massive than the Galilean moons). For the observations, we assume a diffraction-limited telescope with 6.5m diameter circular aperture equivalent collecting area analogous to the planned Habitable Worlds Observatory (HWO, \citealt{feinberg_habitable_2024}). This combination of planet and telescope properties results in astrometric precision of 2.39 mas in observations every 10 hours, which is estimated as the performance given an observing wavelength of 630 nm and contrast-limited performance of 3.33$\times10^{-9}$ in 10 hours (SNR=5 and assuming radii of 1 R$_\odot$ for the star, 1 R$_{Jup}$ for the planet, and albedo of 0.3, with astrometric precision scaling as FWHM/SNR). We assume a three year campaign with two months of the year in which the target is unobservable.

\subsection{Planet Orbit Fitting}

We then fit the orbit of the planet. Generally, imaged exoplanets have periods $\gtrsim$10$-$100 yr, and thus their orbital constraints are limited by the lack of coverage of a full orbit. In contrast, the planets considered here have orbital periods on the order of $\sim$yrs, and thus a complete orbit can be observed within a several year campaign. Therefore, relatively simple methods of orbit fitting are possible, which greatly reduces the computational burden of Markov Chain Monte Carlo methods that are typically used for fitting imaged exoplanet orbits (e.g., \citealt{Blunt2020}). While these more sophisticated methods can be readily incorporated in the future, a computationally inexpensive approach is preferred here, and is justified by the relatively complete orbital coverage possible for habitable-zone planets. It is also justified by our desire for a best-fit orbit of the planet, which can be obtained faster than confidence intervals on the orbital parameters, for which other methods are superior.

Specifically, we use the Newton-Raphson (NR) method developed for astrometric orbit fitting of visual binary stars \citep{Schaefer2016}, which minimizes $\chi^2$ by performing a first-order Taylor expansion for the equations of orbital motion. We begin with guesses for the planetary orbital parameters that are within 5\% of their true value to speed up computation time further. This choice is again realistic and justified, given the nearly complete coverage of the planet's orbit. Nevertheless, the method fails to converge on a local $\chi^2$ minimum in a reasonable amount of time for a fraction of cases, depending on the initial parametric guesses. To mitigate such waste of resources, if the NR method fails to converge within 25 iterations, the procedure is restarted with a different set of initial guesses chosen in the same manner. This results in a best-fit orbit to be obtained for the planet typically in under a minute on a modern desktop computer, even for several thousand simulated measurements, whereas other methods can take hours to complete. In practice, for just a single time series of astrometric measurements, the orbit fitting methodology could be completely replaced with MCMC methods (e.g., \citealt{Blunt2020}) and would only need to be computed once, rather than once per simulation as is required for this current work. 

\subsection{Analysis of Orbital Residuals for Signals of Exomoons}

Once the planet's orbit is determined, we use the retrieved parameters to generate expected positions of a Keplerian orbit at the times of the observations. We plot those over the simulated (true) values for the planet and moon center of mass, true planet location (including its orbit about the planet-moon center of mass), and observed location (including also the simulated noise). The first 25 simulated observations, as well as the predicted locations of the planet, its true position, and position of planet-moon center of mass, are shown in Fig. \ref{orbit}. We note that the fitted planetary orbits tend to most closely match the orbit of the planet-moon center of mass about the star, as expected, rather than the planet's true location or its observed location. At this stage we convert cartesian astrometry to polar coordinates, and focus exclusively on the separation (radial) component. We do not investigate the azimuthal residuals further. This choice is made due to the fact that the planet-to-star separation is a purely relative measurement, whereas the azimuthal position measurement depends also on the ability to determine the absolute orientation of the imaging system, which increases uncertainty. The expected separation of the planet is then subtracted from its observed separation to generate a time-series of separation residuals, in which we search for the exomoon signal.


Common methods for finding periodic signals include Lomb-Scargle periodogram analysis, Fast Fourier Transform methods, and grid searches. For its simplicity, we perform a grid search. Over a grid of 4 to 50 days, with step size of 0.1 days, we folded the time-series of separation residuals and fit a sine function. We then binned the data and selected the period that maximized the difference between the $\chi^2$ of the flat-line fit to the phase folded residuals and the $\chi^2$ of the sine-fit ($\Delta \chi^2$) as a means of selecting the most significant periodic signal. We also compute the error on the retrieved period and amplitude from those that were input. This approach returns the synodic period (i.e., with reference to the star that's being orbited by the planet-moon system), which we convert back to sidereal period for comparison with the input period. We note that $\Delta \chi^2$ can be linked to false alarm probabilities under certain assumptions (if the models are nested and the errors are Gaussian). Instead of false alarm probability, we focus on whether the peak in $\Delta \chi^2$ corresponds to the input period and amplitude, and note that this is typically achieved for $\Delta \chi^2\gtrsim$3 (see Fig. \ref{heatmap}).

\section{Results}

\begin{figure*}[htpb]
\centering
\includegraphics[width=\textwidth]{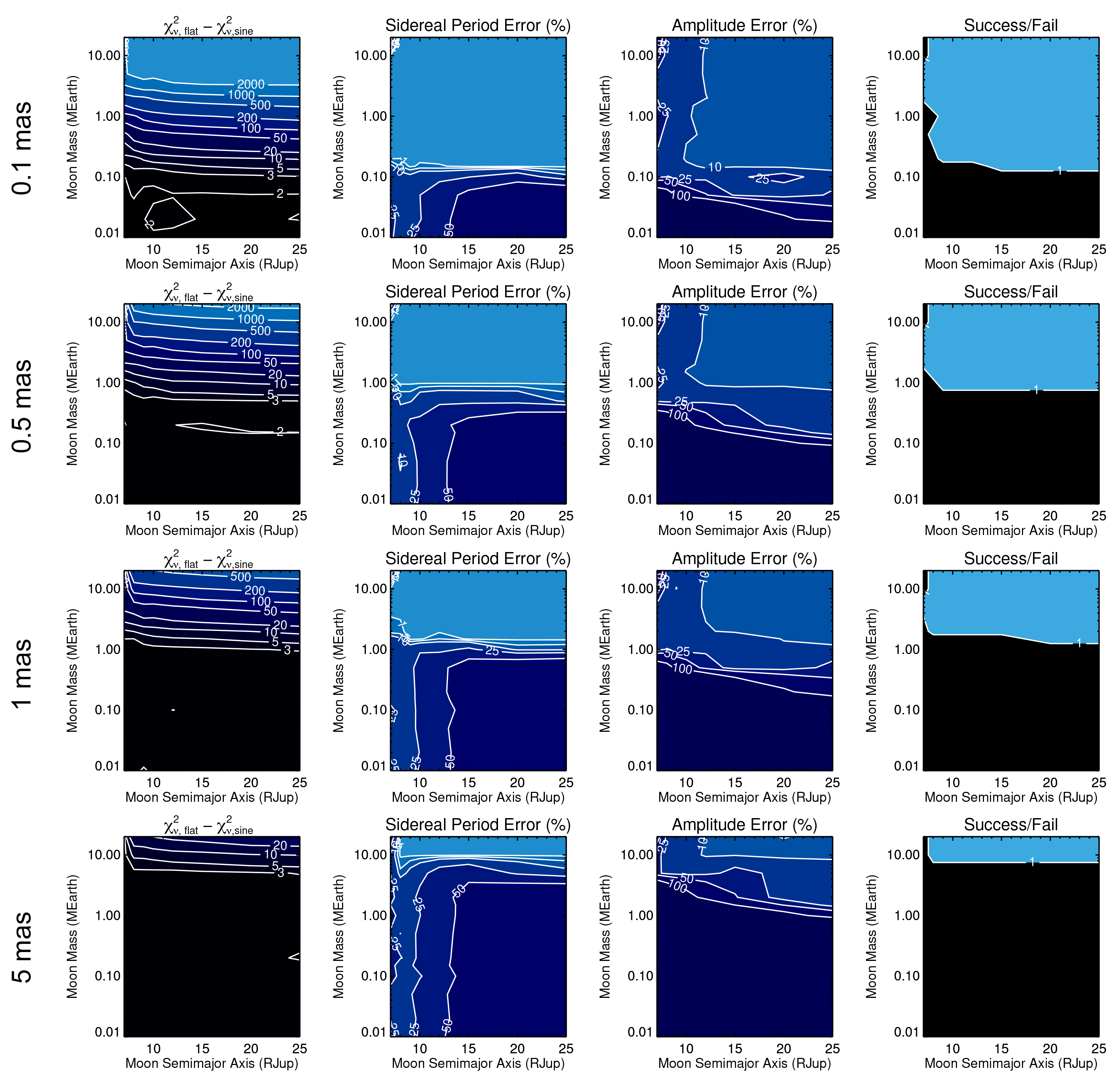}
\caption{\footnotesize \label{heatmap} \textit{Left Column:} $\chi^2$ difference for moon mass ($M_{mn}$) vs. moon orbital semi-major axis ($a_{mn}$) for observing cadence = 1 hr, with no gaps between observations, and a two-month non-observability window per year for a five year campaign.  \textit{Center Columns:} Sidereal period error and  Semi-amplitude error for $M_{moom}$ vs. $a_{mn}$. Rows correspond to varying cases of astrometric precision. \textit{Right Column:} success (blue) defined as the exomoon candidate being recovered with $\chi^2$ difference $>$ 5, period error $<$5\%, and amplitude error $<$25\%. All results are averaged over 20 runs for each combination of $M_{mn}$ vs. $a_{mn}$.}
\end{figure*}

\begin{figure*}[htpb]
\centering
\includegraphics[width=\textwidth]{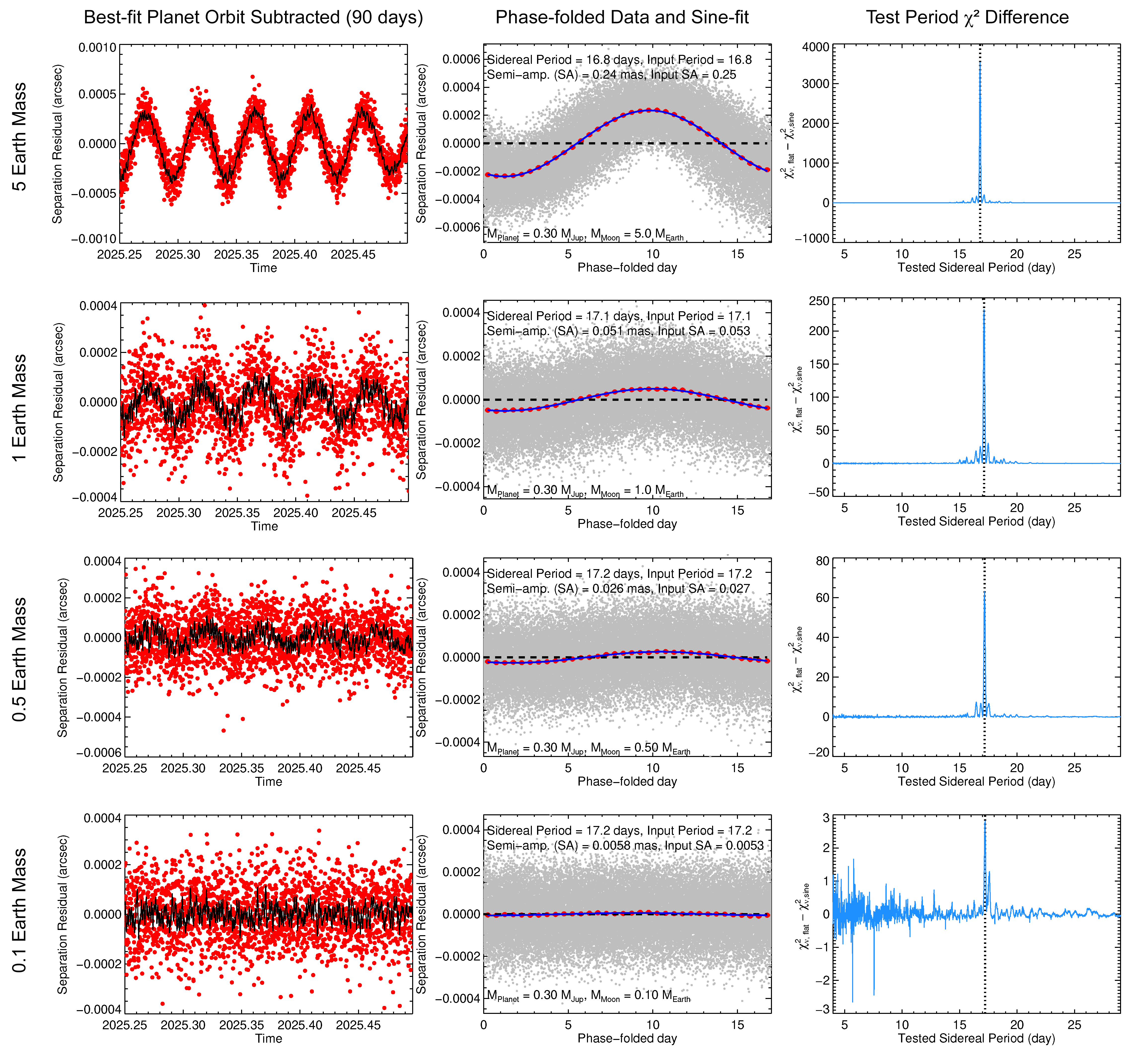}
\caption{\footnotesize \label{masses} Results for astrometric precision = 0.1 mas, observing cadence of 1 hr, and a five year observing campaign for various moon masses assuming a 0.3 M$_{Jup}$ planet orbiting $\alpha$ Centauri A. \textit{Left Column}: first 90 days of the observing campaign, with best-fit planet orbit subtracted (red points,  black curve: smoothed by a running mean of 10 points). \textit{Center Column}: phase-folded data for the full five year campaign corresponding to the largest $\chi^2$ difference from a period grid-search. \textit{Right Column}: $\chi^2$ difference for each of the tested periods. The dashed line represents the known sidereal period of the simulated moon.}

\end{figure*}

\label{results}
In this section we apply the simulation methodology described in $\S$\ref{simulation} to explore the range of exomoon parameters that could plausibly be detectable for the giant planet candidate around $\alpha$ Centuari A \citep{Wagner2021,Beichman2025,Sanghi2025}. Our goal is to demonstrate what would be feasible for a given observatory, rather than to perform a full parameter study of this complex space. We simplify the telescope capabilities to a single parameter $-$ the achievable astrometric precision in a given amount of observing time. We parameterize observing campaign design through the observing cadence and seasonal observability.

The orbital parameters of the planet and moon constitute a fourteen-dimensional parameters space (to which parameters describing the planet and moon brightness could also be added). Typically, such a space would be explored by Monte Carlo methods. However, in this case we are not generalizing to a singular parameter such as goodness of fit. We are instead jointly considering the $\chi^2$ difference, the errors in moon orbital period, and the errors in astrometric semi-amplitude, in order to assess the methodology's effectiveness at recovering the exomoon signals in the simulated datasets. We allow basic properties of the moon (mass and orbital period) to vary, and fix properties of planet mass = 0.3 M$_{Jup}$, planet orbital semi-major axis = 1.8 au, planet orbital inclination = 45$^\circ$, planet orbital eccentricity = 0.3, moon orbital eccentricity = 0.05, and moon orbital inclination = 50$^\circ$.  Aside from the varied parameters, these are identical to those used in the examples in \S \ref{simulation}. The tested grid of $M_{mn}$ vs. $a_{mn}$ in Fig. \ref{heatmap} is (0.01, 0.05, 0.1, 0.15, 0.2, 0.5, 1, 1.5, 2, 5, 10, 20) M$_{\oplus}$~$\times$ (7, 8, 9, 10, 12, 15, 20, 25) $R_{Jup}$. For the assumed planet mass of 0.3 $M{Jup}$, these values of $a_{mn}$ correspond to orbital periods of $T_{mn}$=4.2-28.2 days. The full set simulation parameters is listed in Table \ref{params}. 


\begin{figure*}[htpb]
\centering
\includegraphics[width=\textwidth]{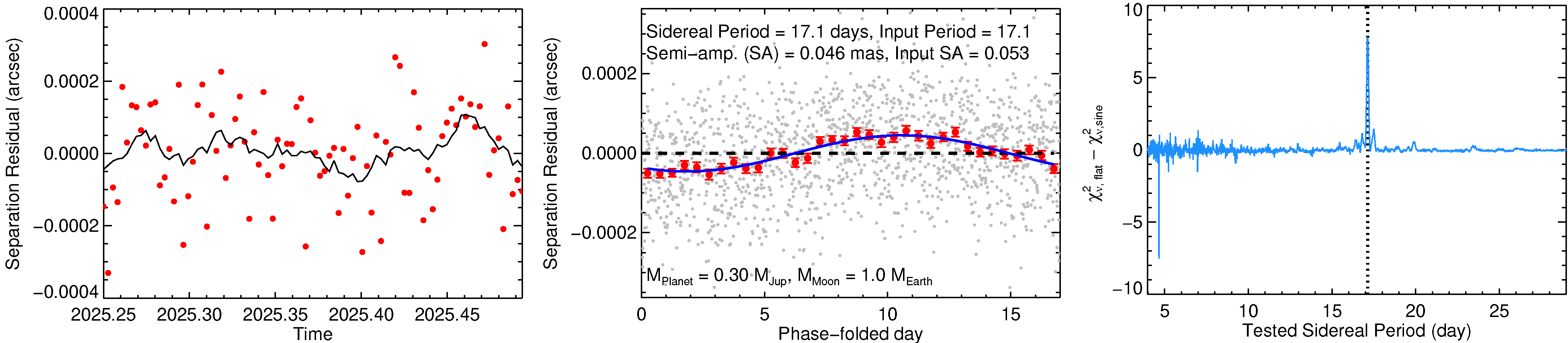}
\caption{\footnotesize \label{elt} Results for astrometric precision = 0.1 mas, observing cadence of 1 day, and a five year observing campaign for an Earth-mass moon orbiting a 0.3 M$_{Jup}$ planet orbiting $\alpha$ Centauri A. \textit{Left}: first 90 days of the observing campaign, with best-fit planet orbit subtracted (red points, smoothed by 10 observations $-$ black curve). \textit{Center}: phase-folded data for the full five year campaign corresponding to the largest $\chi^2$ difference from a period grid-search. \textit{Right}: $\chi^2$ difference for each of the tested periods. The dashed line represents the known sidereal period of the simulated moon.}

\end{figure*}

We fix the system distance to 1.3 pc, motivated by the giant planet candidate orbiting $\alpha$ Centauri A \citep{Wagner2021,Beichman2025,Sanghi2025}. Dependence on system distance, as well as the other fixed parameters, are left for a future parameter study. For the period recovery grid search, we used 4 to 30 days, with step size of 0.01 days, and final binning of 0.1 day intervals. We considered four cases of astrometric precision: 0.1, 0.5, 1, and 5 milli-arcsecond (mas) precision. For all cases, we assume that this level of precision can be obtained in hour-long observations, including overheads, and that the observatory's sole purpose is to monitor the planet's position, except during unobservable windows of $\sim$2 months/yr in which $\alpha$ Centauri would be too close to the Sun for a telescope in an Earth-like orbit. We assume a five year campaign, and that all observations are used in the final analysis. The results are displayed in Fig. \ref{heatmap}. We consider a successful detection to be one in which the sidereal period is recovered to within 5\% error and astrometric amplitude is recovered to within 25\% error, and with $\Delta \chi^2$ $>$5. We permit a larger \% error in amplitude compared to period, because the amplitudes considered are typically sub-mas$-$i.e., smaller than or comparable to the instrumental precision. Moon orbital periods considered are typically on the order of weeks, for which the observing cadence of 1 hr enables sub-percent precision. 

We find that exomoons similar to reported candidates from transit photometry around Kepler-1625 and Kepler 1708 (with masses of $\sim$2.6-10 M$_{\oplus}$: \citealt{Teachey2018b,Kipping2022}) could be detectable with high-confidence even with the poorest astrometric precision tested here (5 mas), highlighting the relative sensitivity of the method. Indeed, exomoons down to Earth-masses (0.2 Earth-masses), are successfully detected across a range of semi-major axes of 5$-$25 R$_{Jup}$ for astrometric precision of 1 mas (0.1 mas), shown in Figs. \ref{heatmap} \& \ref{masses}. For reference, the expected astrometric measurement precision (FWHM/SNR) for SNR=50 and telescope diameter, D=(2.4, 6.5, 25)m are (1.1, 0.41, and 0.11) mas, respectively at 630 nm, and (3.5, 1.3, and 0.34) mas at 2 $\mu$m $-$ representative of the tested range. See \S\ref{existing} for more detailed comparisons to observatory architectures.

As a final example, we explore changing the observing cadence, as the above results have assumed a relatively high cadence of 1 hr, which even if possible, would likely be unfeasible on a public observatory. We consider a practical example of a telescope with D=39 m and observing wavelength of 2 $\mu$m. This configuration would have a diffraction limited FWHM of $\sim$0.06 arcsec. Thus, a SNR$\sim$50 detection would yield astrometric precision of $\sim$0.12 mas. We simulated such observations with a cadence of one ($\sim$hr long) observation per day for five years, with two month/yr non-observability, and found that Earth-mass exomoons could be detected with $\chi^2$ difference $\sim$8 and period and amplitude errors $<$1\% and $\sim$15\%, respectively. The results are shown in Fig. \ref{elt}.




\newpage
\section{Discussion}
\label{discuss}

In \S\ref{simulation}-\ref{results}, we presented a method for extracting the signal of an exomoon from a time-series of relative astrometry between a star and planet. We demonstrated that astrometric precision of $\gtrsim$0.1 mas and observing cadence of $\gtrsim$1 hr can yield high-confidence ($\chi^2_{\nu}$ difference $\gtrsim$ 5) detections of exomoons of $\gtrsim$0.2 M$_{\oplus}$ for a Saturn-mass planet within $\alpha$ Centauri A's habitable zone. Here, we discuss those results and their potential for future applications. 


~
~

\subsection{Physical Assumptions and Limitations}
\label{limits}
Here, we briefly discuss the physical limitations of the models presented above. To start, we only considered systems of one planet, and of one moon orbiting that planet. In reality, there could be multiple planets, and multiple moons, as we have in the solar system. In the case of multiple planets, the orbits of both planets could be modeled simultaneously using the methods presented here. This would allow for moons to be detected around either planet, if both are detected via imaging. If only one planet is detected via imaging, other methods (e.g., radial velocity observations of the star) may be necessary to rule out planets orbiting the star that could be confused with moons. In the case of multiple moons orbiting the same planet, the astrometric signal will be a superposition of the effects of the moons. Using the same methods presented herein, a multi-moon model could be constructed. However, pulling the multiple frequencies out via a grid search would be more complicated than presented, though computationally feasible. Exploring more complex systems than a single planet and moon will be a subject of a future work in this series. 

Additionally, we assumed that the astrometric error budget is dominated by measurement errors. In practice, systematic errors (e.g., determining the position of the star behind the coronagraph) will need to be accounted for. For existing ground-based coronagraphic systems, these are on the order of $\sim$1 mas \citep{Maire2021}. Smearing due to orbital motion of the planet and moon is another potential source of astrometric noise. For the planetary orbit considered in this case study ($a_{pl}$=1.8 au, d=1.3 pc, M$_\star$=1.0 M$_{\odot}$), the planet-moon center of mass (CoM) motion about the star in 10 hrs is $\sim$4 mas. The planet motion about the planet-moon CoM in 10 hrs is $\sim$0.3 mas (for the example in Figs. \ref{orbit} \& \ref{residuals}). As these levels of smearing are similar to the astrometric signal of the moon, smearing could be a complicating factor for long ($\gtrsim$10 hrs) exposures, and generally motivates shorter exposure times for searching for exomoons. However, the individual image exposure times can be much shorter ($\sim$seconds to minutes), and thus once the planet’s orbit has been determined, these individual exposures can be shifted accordingly to account for orbital motion over $\sim$hours in order to eliminate smearing due to the planet+moon CoM about the star. Smearing due to the planet’s motion about the planet+moon CoM is generally smaller, but also motivates shorter exposures. For 1 hr exposures, the effect is $<$0.1 mas for the example in Figs. \ref{orbit} \& \ref{residuals} of a 30 $M_\oplus$ moon with a 32 day period about a Saturn-mass planet. 

We assumed that the planet is observable throughout the seasonal observability window. In reality, there may be periods of time in which the planet is too close to the star to be observed by the coronagraphic imaging system (as is the case for the candidate planet around $\alpha$ Cen A as observed by JWST: \citealt{Sanghi2025}). In such cases, campaigns may need to be extended in order to achieve the same level of detection limits. As a final note, we did not consider the possibility of ringed planets. In practice, such rings could raise the amount of reflected light from the planet by up to a factor of $\sim$4 for Saturn-like rings, improving the achievable level of SNR in a given amount of exposure time. However, rings can also add noise through the photoshift due to the motion of the planet's shadow. This effect will proceed on timescales proportional to the planet’s orbit, which allows for the photocenter shift due to shadowed rings to be separated from the signal of the moons, and possibly even completely removed from the astrometric residuals with a high-pass filter. For Saturn-like rings, in which the main bright rings extend to a few times the planet’s radius, at d=1.3pc would have a radius of $\sim$1 mas. Assuming that a Saturn-diameter shadow is cast on the rings by the planet, the maximum photocenter shift would be smaller than this by an order unity factor, depending on inclination and thickness of the rings.  

\subsection{Comparison to Other Methods}
\label{compare}

These methods for detecting exomoons through relative astrometry between a star and planet are complementary to other existing and proposed methods for exploring the galactic population of exomoons. Compared to transit-based methods (e.g., \citealt{Simon2007,Kipping2009,Teachey2018b,Kipping2022}), those presented herein are relatively limited to nearby stars (though comparable to radio emission: \citealt{Noyola2014}), and relatively expensive. As the astrometric signal scales inversely with distance, even stars at 10 pc are $\sim$7.7$\times$ more challenging than the example of $\alpha$ Centauri. However, the astrometric methods presented herein are (in principle) capable of detecting moons as small as $\sim$0.2 M$_{\oplus}$ at d=1.3 pc, which is $\sim$13 times smaller than the candidate reported in \cite{Kipping2022} at 4.8-$\sigma$. Of course, this comes at significantly greater cost per target. For instance, The Kepler mission \citep{Borucki2010} had a primary mirror diameter of 1.4m, operated for $\sim$4 years with a cadence of 30 min, and observed thousands of stars out to kiloparsec distances. A space telescope dedicated to searching for exomoons would have a primary mirror diameter of $\sim$3-6m, a cadence of $\sim$minutes (for individual exposures), and would only observe a handful of targets (perhaps just a single target) for a similar duration of $\sim$3-5 years. Such a mission is conceivable given the rapid advancements in launch technology that are driving down mission costs (e.g., \citealt{Ansar2022, Douglas2023}).

Nearby stars offer several other advantages$-$not least of which include possibilities for robotic exploration (e.g., \citealt{Parker2018}). Such nearby moons are also potentially interesting targets for follow-up characterization, as any light from the moon itself will also be brighter by a factor of the square of their ratio of distances (all else being equal). As for such direct methods of detection, spectroastrometry (e.g., \citealt{Agol2015}) is perhaps the most promising for moons without an additional source of heating (such as tides, e.g., \citealt{Limbach2013}). In practice, spectroscopy can be combined with the methods presented here if the observing instrument is an integral field spectrograph, like many existing exoplanet imaging instruments (e.g., \citealt{Macintosh2014, Stone2018,Beuzit2019,Sallum2023,Ruffio2024,Wright2023}). Careful consideration of spectral resolution combined with a sufficient spectral signal could allow spectral confirmation and characterization, combined with dynamical mass measurement via astrometry, in a single mode of observation. We note that the amount of observing time for either method (relative astrometric time-series, or deep integrated spectroscopy) would be significant for even a low SNR detection of an exomoon, and thus simultaneous confirmation from multiple methods would be highly beneficial, and would enable more complete constraints on the exomoon's properties. 

\subsection{Prospects for Existing and Future Telescopes}
\label{existing}

The largest ground-based telescope under construction is the D=39 m Extremely Large Telescope (ELT). As discussed in \S\ref{results}, such a system would have a diffraction limited FWHM of $\sim$0$\farcs$06 arcsec. Thus, to reach $\sim$0.1 mas astrometric prescision requires SNR$\sim$60. A first-generation instrument, METIS \citep{Brandl2021}, is expected to operate between 3-13 $\mu$m, while a proposed second-generation instrument, PCS \citep{Kasper2021} aims to combine high-resolution spectroscopy with coronagraphy at shorter wavelengths. Both instruments aim to detect habitable-zone Earth-sized planets with SNR$\gtrsim$5 in $\lesssim$1 night of observations. Extrapolating from 1 R$_{\oplus}$ to 1 R$_{Jup}$ results in giant planets being two orders of magnitude brighter than Earth-sized planets, and thus SNR$\gtrsim$100 detections of habitable-zone giant planets should be feasible within a single night of observing. If such observations can be completed in a matter of hours while enabling concurrent science operations, it is feasible that the ELT could provide nightly coverage of the position of a habitable-zone giant planet orbiting $\alpha$ Centauri A with $\sim$0.1 mas astrometric precision. In \S\ref{results}, we showed that a five year campaign with such a system would be sufficient for the detection of an Earth-mass moon, assuming a Saturn-mass planet (consistent with the radial velocity detection limits: \citealt{Zhao2018}). An Earth-mass moon to a Saturn-mass planet would have a mass ratio of $q\sim$ 0.01, which is similar to the mass ratio of Earth and its moon ($q\sim$  0.012). 

The Habitable Worlds Observatory (HWO) concept \citep{astro2020} is also expected to achieve SNR$\gtrsim$5 detections of Earth-sized planets in $\sim$hour long exposures out to d$\sim$10pc. Assuming D=6 m and observing wavelength of 500 nm, HWO would have a FWHM $\sim$ 0$\farcs$02. Thus, it will also achieve SNR$\gtrsim$100 and astrometric precision of $\lesssim$0.1 mas on habitable-zone giant planets on timescales of hours. However, at d=10pc the astrometric signal from a hypothetical exomoon would also be reduced by a factor of 10. Nevertheless, super-Earth-mass moons would still be detectable orbiting Saturn-mass planets with one $\sim$hour long observation per day even for these more distant targets. For systems in which HWO is already spending considerable amounts of telescope time to acquire spectroscopy of rocky planets, searching for exomoons around giant planets that are simultaneously observable (i.e., within the outer working angle) in the system could be accomplished at no added cost. In this case, the detectable moon period range could be enlarged by carefully planned visit timing (e.g., one observation per day for several years, rather than one continuous several hundred hour long observation). 

\subsection{Application to the Planet Candidate $\alpha$ Centauri Ab $-$ Motivation for a Dedicated Space Telescope}
\label{future}

Finally, this study was motivated by the giant planet candidate reported around $\alpha$ Centauri A \citep{Wagner2021,Beichman2025,Sanghi2025}, $\alpha$ Cen Ab. Being the closest G-type star, such a configuration would be optimally suited for the astrometric detection of exomoons. The signal from a moon orbiting an identical planet around the next closest G-type star would be $\sim$2.7$\times$ smaller than for $\alpha$ Centauri, requiring a much more capable observatory to achieve similar exomoon detection limits. Given that habitable-zone exomoons could also plausibly support life (e.g., \citealt{Reynolds1987,Williams1997,Scharf2006}), the candidate planet $\alpha$~Cen~Ab could host some of the nearest habitable environments to the Sun. Using 0.1 mas as a reasonable estimate of astrometric precision that can be attained, the results in \S\ref{results} show that $\sim$0.2 M$_{\oplus}$ moons with $a_{mn}$=5-25 R$_{Jup}$ could be detected in a five year campaign with a dedicated telescope. Such moons would be stable to the gravitational influence of $\alpha$ Centauri B \citep{RosarioFranco2020,Quarles2021}, and would have mass ratios of $q\sim$10$^{-3}$, of which there are examples in the Solar System ($q_{Titan-Saturn}\sim2\times 10^{-4}$, $q_{Moon-Earth}\sim10^{-2}$). Therefore, such moons could plausibly exist. Should this candidate planet be confirmed, a dedicated space observatory for monitoring the planet for exomoons could become a reasonably compelling undertaking. 

A 3m-class telescope could be conceivable as a purpose-built, near-term option for searching for exomoons within the $\alpha$ Centauri system (e.g., \citealt{Douglas2023}). Such a telescope would have a diffraction-limited FWHM of $\sim$0$\farcs$035 at 500 nm. To achieve astrometric precision of $\sim$1 mas, sufficient to detect $\sim$Earth-mass moons, would require SNR $\sim$35 in $\sim$1 hr, assuming a Saturn-mass planet and a five-year campaign on a dedicated telescope. A Saturn-radius planet with albedo = 0.2 at $a_{pl}$=1.8 au from $\alpha$ Cen A would have a maximum contrast ratio of $\sim$5$\times$10$^{-9}$. Reaching the required SNR would thus require SNR=5 contrast ratios of $\sim$10$^{-9}$, which recent lab results have shown is feasible \citep{Anche2024}.


\section{Conclusion}

\label{summary}


In summary, we developed a simulation framework for detecting exomoons via relative astrometry of directly imaged giant exoplanets. This framework simulates the physical position of a planet orbiting a star, that itself is orbited by a moon. It includes noise, and then attempts to find the signal of the moon and recover its parameters in the simulated dataset. We focus this simulation framework on the $\alpha$ Centauri system, motivated by the giant planet candidate \citep{Wagner2021,Beichman2025,Sanghi2025}. We explore parameters of moon to planet mass ratio, moon semi-major axis, observing cadence, and astrometric precision, while fixing the other parameters to those consistent with the planet candidate (see Table \ref{params} and \citealt{Beichman2025}). 

We find that detectability of moons improves with more frequent observing cadence and with a greater number of observations, as expected for a periodic signal near the instrumental precision within a timeseries dataset. We explored applications of these results to ground-based telescopes under construction, such as the 39m Extremely Large Telescope. We find that Earth-mass moons could be detectable around the $\alpha$ Centauri candidate planet with astrometric precision of 0.1 mas with 24 hr cadence over a 5 year campaign. We also explored application of these results to HWO, finding that moons of several Earth-masses orbiting giant planets out to $d=10$pc can be detected with cadence of 24 hr in a 5 year campaign. This could possibly be done concurrently to measuring spectra of other planets in the system. Finally, we discussed what could be accomplished with a dedicated space telescope. We find that moons with mass of $\sim$0.2 M$_{\oplus}$ can be detected at separations of 10-25 R$_{Jup}$ for astrometric precision of 0.1 mas and cadence of 1 hr over a five year campaign (assuming a Saturn mass planet). These limits are within the range of moon to planet mass ratios found in the solar system. 

We conclude that exomoons of Earth-mass or smaller could realistically be detectable around $\alpha$ Cen Ab within the next decade, given sufficient precision ($\sim$mas) and observing coverage ($\sim$5yr duration, $\sim$1 hr cadence). This would provide motivation for a dedicated space telescope, should the planet candidate be confirmed. Generally, these results support the development of high-precision astrometric monitoring campaigns as a viable path to detecting habitable-zone exomoons around nearby stars.




\section{Acknowledgments} KW acknowledges support from The Breakthrough Prize Foundation. The results reported herein benefited from collaborations and/or information exchange within NASA's Nexus for Exoplanet System Science (NExSS) research coordination network sponsored by NASA's Science Mission Directorate. This material is based on work supported by the National Science Foundation Graduate Research Fellowship under Grant No.~2139433.


\begin{thebibliography}{}

\bibitem[Agol et al.(2015)]{Agol2015} Agol, E., Jansen, T., Lacy, B., et al.\ 2015, \apj, The Center of Light: Spectroastrometric Detection of Exomoons, 812, 1, 5. doi:10.1088/0004-637X/812/1/5

\bibitem[Anche et al.(2024)]{Anche2024} Anche, R.~M., Van Gorkom, K.~J., Ashcraft, J.~N., et al.\ 2024, \procspie, 13092, 130926M. doi:10.1117/12.3020630

\bibitem[Ansar \& Flyvbjerg(2022)]{Ansar2022} Ansar, A. \& Flyvbjerg, B.\ 2022, , arXiv:2206.08754. doi:10.48550/arXiv.2206.08754

\bibitem[Beichman et al.(2020)]{Beichman2019} Beichman, C., Ygouf, M., Llop Sayson, J., et al.\ 2020, \pasp, Searching for Planets Orbiting {\ensuremath{\alpha}} Cen A with the James Webb Space Telescope, 132, 1007, 015002. doi:10.1088/1538-3873/ab5066

\bibitem[Beichman et al.(2025)]{Beichman2025} Beichman, C., et al.\ 2025, \apjl, in press

\bibitem[Bennett et al.(2014)]{Bennet2014} Bennett, D.~P., Batista, V., Bond, I.~A., et al.\ 2014, \apj, MOA-2011-BLG-262Lb: A Sub-Earth-Mass Moon Orbiting a Gas Giant Primary or a High Velocity Planetary System in the Galactic Bulge, 785, 2, 155. doi:10.1088/0004-637X/785/2/155

\bibitem[Beuzit et al.(2019)]{Beuzit2019} Beuzit, J.-L., Vigan, A., Mouillet, D., et al.\ 2019, \aap, 631, A155. doi:10.1051/0004-6361/201935251

\bibitem[Borucki et al.(2010)]{Borucki2010} Borucki, W.~J., Koch, D., Basri, G., et al.\ 2010, Science, 327, 5968, 977. doi:10.1126/science.1185402

\bibitem[Blunt et al.(2020)]{Blunt2020} Blunt, S., Wang, J.~J., Angelo, I., et al.\ 2020, \aj, orbitize!: A Comprehensive Orbit-fitting Software Package for the High-contrast Imaging Community, 159, 3, 89. doi:10.3847/1538-3881/ab6663

\bibitem[Bond et al.(2015)]{Bond2015} Bond, H.~E., Gilliland, R.~L., Schaefer, G.~H., et al.\ 2015, \apj, Hubble Space Telescope Astrometry of the Procyon System, 813, 2, 106. doi:10.1088/0004-637X/813/2/106

\bibitem[Brandl et al.(2021)]{Brandl2021} Brandl, B., Bettonvil, F., van Boekel, R., et al.\ 2021, The Messenger, 182, 22. doi:10.18727/0722-6691/5218


\bibitem[Brandt(2021)]{Brandt2021} Brandt, T.~D.\ 2021, \apjs, The Hipparcos-Gaia Catalog of Accelerations: Gaia EDR3 Edition, 254, 2, 42. doi:10.3847/1538-4365/abf93c

\bibitem[Cameron(2009)]{Cameron2009} Cameron, J. 2009, \textit{Avatar}, 20th Century Fox, USA 

\bibitem[Cilibrasi et al.(2021)]{Calibrisi2021} Cilibrasi, M., Szul{\'a}gyi, J., Grimm, S.~L., et al.\ 2021, \mnras, 504, 4, 5455. doi:10.1093/mnras/stab1179


\bibitem[Currie et al.(2023)]{Currie2023} Currie, T., Brandt, G.~M., Brandt, T.~D., et al.\ 2023, Science, Direct imaging and astrometric detection of a gas giant planet orbiting an accelerating star, 380, 6641, 198. doi:10.1126/science.abo6192


\bibitem[Douglas et al.(2023)]{Douglas2023} Douglas, E.~S., Aldering, G., Allan, G.~W., et al.\ 2023, \procspie, 12677, 126770D. doi:10.1117/12.2677843

\bibitem[Feinberg et al (2024)]{feinberg_habitable_2024} Feinberg, Lee and Ziemer, John and Ansdell, Megan and Crooke, Julie and Dressing, Courtney and Mennesson, Bertrand and O'Meara, John and Pepper, Joshua and Roberge, Aki.\ 2025, Proc SPIE.
  doi:10.1117/12.3018328,


\bibitem[Foust et al.(1997)]{Foust1997} Foust, J.~A., Elliot, J.~L., Olkin, C.~B., et al.\ 1997, \icarus, Determination of the Charon/Pluto Mass Ratio from Center-of-Light Astrometry, 126, 2, 362. doi:10.1006/icar.1996.5656

\bibitem[Franson et al.(2023)]{Franson2023} Franson, K., Bowler, B.~P., Zhou, Y., et al.\ 2023, \apjl, 950, 2, L19. doi:10.3847/2041-8213/acd6f6


\bibitem[Galilei(1610)]{Galilei1610} Galilei, G.\ 1610, Sidereus nuncius magna, longeque admirabilia spectacula pandens lunae facie, fixis innumeris, lacteo circulo, stellis nebulosis, ... Galileo Galileo : nuper a se reperti beneficio sunt observata in apprime vero in quatuor planetis circa Iovis stellam disparibus intervallis, atque periodis, celeritate mirabili circumvolutis ... atque Medicea sidera nuncupandos decrevit. doi:10.3931/e-rara-695

\bibitem[Heller(2016)]{Heller2016} Heller, R.\ 2016, \aap, 588, A34. doi:10.1051/0004-6361/201527496


\bibitem[Heller et al.(2019)]{Heller2019} Heller, R., Rodenbeck, K., \& Bruno, G.\ 2019, \aap, An alternative interpretation of the exomoon candidate signal in the combined Kepler and Hubble data of Kepler-1625, 624, A95. doi:10.1051/0004-6361/201834913

\bibitem[Heller \& Hippke(2024)]{Heller2024} Heller, R. \& Hippke, M.\ 2024, Nature Astronomy, Large exomoons unlikely around Kepler-1625 b and Kepler-1708 b, 8, 193. doi:10.1038/s41550-023-02148-w


\bibitem[Kasper et al.(2021)]{Kasper2021} Kasper, M., Cerpa Urra, N., Pathak, P., et al.\ 2021, The Messenger, 182, 38. doi:10.18727/0722-6691/5221

\bibitem[Kipping et al.(2009)]{Kipping2009} Kipping, D.~M., Fossey, S.~J., \& Campanella, G.\ 2009, \mnras, On the detectability of habitable exomoons with Kepler-class photometry, 400, 1, 398. doi:10.1111/j.1365-2966.2009.15472.x

\bibitem[Kipping et al.(2022)]{Kipping2022} Kipping, D., Bryson, S., Burke, C., et al.\ 2022, Nature Astronomy, An exomoon survey of 70 cool giant exoplanets and the new candidate Kepler-1708 b-i, 6, 367. doi:10.1038/s41550-021-01539-1

\bibitem[Kipping et al.(2024)]{Kipping2025} Kipping, D., Teachey, A., Yahalomi, D.~A., et al.\ 2024, , arXiv:2401.10333. doi:10.48550/arXiv.2401.10333

\bibitem[Kreidberg et al.(2019)]{Kreidberg2019} Kreidberg, L., Luger, R., \& Bedell, M.\ 2019, \apjl, No Evidence for Lunar Transit in New Analysis of Hubble Space Telescope Observations of the Kepler-1625 System, 877, 2, L15. doi:10.3847/2041-8213/ab20c8

\bibitem[Lazzoni et al.(2022)]{Lazzoni2022} Lazzoni, C., Desidera, S., Gratton, R., et al.\ 2022, \mnras, 516, 1, 391. doi:10.1093/mnras/stac2081


\bibitem[Macintosh et al.(2014)]{Macintosh2014} Macintosh, B., Graham, J.~R., Ingraham, P., et al.\ 2014, Proceedings of the National Academy of Science, 111, 35, 12661. doi:10.1073/pnas.1304215111



\bibitem[Maire et al.(2021)]{Maire2021} Maire, A.-L., Langlois, M., Delorme, P., et al.\ 2021, Journal of Astronomical Telescopes, Instruments, and Systems, 7, 035004. doi:10.1117/1.JATIS.7.3.035004

\bibitem[Marius(1614)]{Marius1614} Marius, S.\ 1614, Mundus Iovialis Anno M.DC.IX.Detectus Ope Perspicill Belgici, Hoc est, Quatuor Jovialium Planetarum, cum Theoria, tum Tabulæ, Propriis Observationibus Maxime Fundatæ. 


\bibitem[Michaelis et al.(2025)]{Michaelis2025} Michaelis, M.~B., Lietzow-Sinjen, M., \& Wolf, S.\ 2025, \aap, 696, A208. doi:10.1051/0004-6361/202452870

\bibitem[National Academies of Sciences(2021)]{astro2020} National Academies of Sciences, E.\ 2021, . doi:10.17226/26141

\bibitem[Noyola et al.(2014)]{Noyola2014} Noyola, J.~P., Satyal, S., \& Musielak, Z.~E.\ 2014, \apj, 791, 1, 25. doi:10.1088/0004-637X/791/1/25


\bibitem[Parkin(2018)]{Parker2018} Parkin, K.~L.~G.\ 2018, Acta Astronautica, 152, 370. doi:10.1016/j.actaastro.2018.08.035

\bibitem[Pasachoff(2015)]{Pasachoff2015} Pasachoff, J.~M.\ 2015, Journal for the History of Astronomy, 46, 2, 218. doi:10.1177/0021828615585493


\bibitem[Peters-Limbach \& Turner(2013)]{Limbach2013} Peters-Limbach, M.~A. \& Turner, E.~L.\ 2013, \apj, On the Direct Imaging of Tidally Heated Exomoons, 769, 2, 98. doi:10.1088/0004-637X/769/2/98


\bibitem[Quarles et al.(2021)]{Quarles2021} Quarles, B., Eggl, S., Rosario-Franco, M., et al.\ 2021, \aj, Exomoons in Systems with a Strong Perturber: Applications to {\ensuremath{\alpha}} Cen AB, 162, 2, 58. doi:10.3847/1538-3881/ac042a


\bibitem[Rosario-Franco et al.(2020)]{RosarioFranco2020} Rosario-Franco, M., Quarles, B., Musielak, Z.~E., et al.\ 2020, \aj, Orbital Stability of Exomoons and Submoons with Applications to Kepler 1625b-I, 159, 6, 260. doi:10.3847/1538-3881/ab89a7

\bibitem[Ruffio et al.(2023)]{Ruffio2023} Ruffio, J.-B., Horstman, K., Mawet, D., et al.\ 2023, \aj, 165, 3, 113. doi:10.3847/1538-3881/acb34a

\bibitem[Ruffio et al.(2024)]{Ruffio2024} Ruffio, J.-B., Perrin, M.~D., Hoch, K.~K.~W., et al.\ 2024, \aj, 168, 2, 73. doi:10.3847/1538-3881/ad5281


\bibitem[Reynolds et al.(1987)]{Reynolds1987} Reynolds, R.~T., McKay, C.~P., \& Kasting, J.~F.\ 1987, Advances in Space Research, 7, 5, 125. doi:10.1016/0273-1177(87)90364-4

\bibitem[Sanghi et al.(2025)]{Sanghi2025} Sanghi, A., et al.\ 2025, \apjl, in press

\bibitem[Schaefer et al.(2016)]{Schaefer2016} Schaefer, G.~H., Hummel, C.~A., Gies, D.~R., et al.\ 2016, \aj, Orbits, Distance, and Stellar Masses of the Massive Triple Star {\ensuremath{\sigma}} Orionis, 152, 6, 213. doi:10.3847/0004-6256/152/6/213

\bibitem[Sallum et al.(2023)]{Sallum2023} Sallum, S., Skemer, A., Stelter, D., et al.\ 2023, , arXiv:2310.07134. doi:10.48550/arXiv.2310.07134

\bibitem[Scharf(2006)]{Scharf2006} Scharf, C.~A.\ 2006, \apj, 648, 2, 1196. doi:10.1086/505256


\bibitem[Simon et al.(2007)]{Simon2007} Simon, A., Szatm{\'a}ry, K., \& Szab{\'o}, G.~M.\ 2007, \aap, Determination of the size, mass, and density of ``exomoons'' from photometric transit timing variations, 470, 2, 727. doi:10.1051/0004-6361:20066560

\bibitem[Stone et al.(2018)]{Stone2018} Stone, J.~M., Skemer, A.~J., Hinz, P., et al.\ 2018, \procspie, 10702, 107023F. doi:10.1117/12.2313977


\bibitem[Teachey et al.(2018)]{Teachey2018a} Teachey, A., Kipping, D.~M., \& Schmitt, A.~R.\ 2018, \aj, HEK. VI. On the Dearth of Galilean Analogs in Kepler, and the Exomoon Candidate Kepler-1625b I, 155, 1, 36. doi:10.3847/1538-3881/aa93f2


\bibitem[Teachey \& Kipping(2018)]{Teachey2018b} Teachey, A. \& Kipping, D.~M.\ 2018, Science Advances, Evidence for a large exomoon orbiting Kepler-1625b, 4, 10, eaav1784. doi:10.1126/sciadv.aav1784

\bibitem[Tuthill et al.(2018)]{Tuthill2018} Tuthill, P., Bendek, E., Guyon, O., et al.\ 2018, \procspie, The TOLIMAN space telescope, 10701, 107011J. doi:10.1117/12.2313269

\bibitem[Vanderburg et al.(2018)]{Vanderburg2018} Vanderburg, A., Rappaport, S.~A., \& Mayo, A.~W.\ 2018, \aj, 156, 5, 184. doi:10.3847/1538-3881/aae0fc



\bibitem[Wagner et al.(2021)]{Wagner2021} Wagner, K., Boehle, A., Pathak, P., et al.\ 2021, Nature Communications, Imaging low-mass planets within the habitable zone of {\ensuremath{\alpha}} Centauri, 12, 922. doi:10.1038/s41467-021-21176-6

\bibitem[Weible et al.(2025)]{Weible2025} Weible, G., Wagner, K., Stone, J., et al.\ 2025, \aj, 169, 4, 197. doi:10.3847/1538-3881/adadf6


\bibitem[Wang et al.(2014)]{Wang2014} Wang, J.~J., Rajan, A., Graham, J.~R., et al.\ 2014, \procspie, 9147, 914755. doi:10.1117/12.2055753

\bibitem[Williams et al.(1997)]{Williams1997} Williams, D.~M., Kasting, J.~F., \& Wade, R.~A.\ 1997, \nat, 385, 6613, 234. doi:10.1038/385234a0


\bibitem[Wright et al.(2023)]{Wright2023} Wright, G.~S., Rieke, G.~H., Glasse, A., et al.\ 2023, \pasp, 135, 1046, 048003. doi:10.1088/1538-3873/acbe66


\bibitem[Zhao et al.(2018)]{Zhao2018} Zhao, L., Fischer, D.~A., Brewer, J., et al.\ 2018, \aj, Planet Detectability in the Alpha Centauri System, 155, 1, 24. doi:10.3847/1538-3881/aa9bea



\end{thebibliography}
\end{document}